\documentclass{myarticle}
\usepackage{amsmath,amsfonts,amssymb}
\usepackage{mathtools}
\usepackage{accents} 
\usepackage{lineno}
\usepackage[hypertexnames=false]{hyperref} 
\usepackage{graphicx}
\usepackage{relsize}
\usepackage[nice]{nicefrac}
\usepackage{comment}

\biboptions{sort&compress}  

\newcommand{\defeq}{\vcentcolon=}

\newcommand{\subs}[1] {_{_{\text{\tiny #1}}}}
\newcommand{\supt}[1] {\,{^{^{\text{\tiny #1}}}}}

\newcommand{\msub}[1] {_{_{\text{\scriptsize $#1$}}}}
\newcommand{\msubs}[1] {_{_{\text{\tiny $#1$}}}}

\newcommand{\nexp}[2]{e^{-\,\nicefrac{\mathlarger{#1}}{\mathlarger{#2}}}}
\newcommand{\eps}{\epsilon}
\newcommand{\suchthat}{\mathlarger{\backepsilon}\,'}
\newcommand{\wflx}{w\msubs{\!f\!l\!x}}
\newcommand{\bhat}[1]{\ensuremath{\accentset{\wedge}{#1}}}
\newcommand{\bcheck}[1]{\ensuremath{\accentset{\vee}{#1}}}
\newcommand{\Fsw}[1]{F\!\supt{(#1)}\![w]}
\newcommand{\fsw}[1]{f\!\supt{(#1)}\![w]}
\newcommand{\wuu}{\prescript{}{}w^{\mathsmaller{(1)}}_{1}} 
\newcommand{\wud}{\prescript{}{}w^{\mathsmaller{(1)}}_{2}}
\newcommand{\wdu}{\prescript{}{}w^{\mathsmaller{(2)}}_{1}}
\newcommand{\wdd}{\prescript{}{}w^{\mathsmaller{(2)}}_{2}}
\newcommand{\wfu}{\prescript{}{}w^{\mathsmaller{(f)}}_{1}}
\newcommand{\wfd}{\prescript{}{}w^{\mathsmaller{(f)}}_{2}}
\newcommand{\wf}{\prescript{}{}w^{\mathsmaller{(f)}}_{}}
\newcommand{\wU}{\prescript{}{}w^{\mathsmaller{(1)}}_{}}
\newcommand{\wD}{\prescript{}{}w^{\mathsmaller{(2)}}_{}}
\newcommand{\wTre}{\prescript{}{}w^{\mathsmaller{(3)}}_{}}
\newcommand{\wT}{\prescript{}{}w^{\mathsmaller{(T)}}_{}}
\newcommand{\wQ}{\prescript{}{}w^{\mathsmaller{(4)}}_{}}
\newcommand{\wUc}{\prescript{}{}{\bcheck{w}}^{\mathsmaller{(1)}}_{}}

\newcommand{\wTc}{\prescript{}{}{\bcheck{w}}^{\mathsmaller{(3)}}_{}}

\newcommand{\bwU}{\prescript{}{}{\bcheck{w}}^{\mathsmaller{(1)}}_{}}
\newcommand{\bwf}{\prescript{}{}{\bcheck{w}}^{\mathsmaller{(f)}}_{}}
\newcommand{\bwD}{\prescript{}{}{\bcheck{w}}^{\mathsmaller{(2)}}_{}}
\newcommand{\bwT}{\prescript{}{}{\bcheck{w}}^{\mathsmaller{(3)}}_{}}
\newcommand{\bwL}{\prescript{}{}{\bcheck{w}}^{\mathsmaller{(T)}}_{}}
\newcommand{\bwQ}{\prescript{}{}{\bcheck{w}}^{\mathsmaller{(4)}}_{}}

\newcommand{\dewu}{\prescript{}{}{\delta w}^{\mathsmaller{(1)}}_{}}
\newcommand{\dewd}{\prescript{}{}{\delta w}^{\mathsmaller{(2)}}_{}}
\newcommand{\dewf}{\prescript{}{}{\delta w}^{\mathsmaller{(f)}}_{}}
\newcommand{\ecirc}[1]{{#1}^{\mathsmaller{\bf \circ}}}
\newcommand{\ecup}[1]{{#1}^{\bf \mathsmaller{u}}}

\begin{document}

\begin{frontmatter}

\title{An analytic approximate solution of the SIR model}

\author[mysecondaryaddress]{I. Lazzizzera\corref{mycorrespondingauthor}}
\address{Associated with Department of Physics - University of Trento - Italy}
\address{Associated with  Trento Institute for Fundamental Physics and Applications - INFN - Italy}
\address{via Sommarive 14 - 38123 Povo (TN) Italy}
\cortext[mycorrespondingauthor]{Corresponding author}
\ead{ignazio.lazzizzera@unitn.it}

\begin{abstract}
  The SIR(D) epidemiological model is defined through a system of transcendental equations, not solvable by elementary functions. In the present paper those equations are successfully replaced by approximate ones, whose solutions are given explicitly in terms of elementary functions, originating, piece-wisely, from generalized logistic functions: they ensure {\em exact} (in the numerical sense) asymptotic values, besides to be quite practical to use, for example with fit to data algorithms; moreover they unveil a useful feature, that in fact, at least with very strict approximation, is also owned by the (numerical) solutions of the {\em exact} equations. The novelties in the work are:  the way the approximate equations are obtained, using simple, analytic geometry considerations; the easy and practical formulation of the final approximate solutions; the mentioned useful feature, never disclosed before. The work's method and result prove to be robust over a range of values of the well known non-dimensional parameter called {\em basic reproduction ratio}, that covers at least all the known epidemic cases, from influenza to measles: this is a point which doesn't appear much discussed in analogous works.
\end{abstract}

\begin{keyword}
\texttt{\\SIR epidemic model, Kermack-McKendrick model, epidemic dynamics, approximate analytic solution.}
\end{keyword}

\end{frontmatter}


\section{Introduction}
The \textbf{SIR model} \cite{Kermack_McKendrick, Murray:1993, Daley_Gani, Brauer:2017, Martcheva, Brauer_Castillo-Chavez_Feng} is a simple {\em compartmental model} of infectious diseases developed by Kermack and McKendrick \cite{Kermack_McKendrick} in 1927. It considers three compartments: \newline
\textbf{S}, the set of susceptible individuals; \newline
\textbf{I}, the set of the infectious (or {\em currently positive}) individuals, who have been infected and are capable of infecting susceptible individuals; \newline
\textbf{R}, the set of the {\em removed} individuals, namely people who recovered ({\em healed}, \textbf{H} subset) from the disease or deceased due to the disease (\textbf{D} subset), the former assumed to remain immune afterwards. \newline
The SIR model does not consider at all the sub-compartments \textbf{H} and \textbf{D}; instead the SIRD model simply assumes them to constitute a partition of \textbf{R}, fractionally fixed over time, so that, actually compared to the \textbf{SIR} model, nothing substantially changes in the dynamics of the epidemic progression. \newline
It is assumed that births and non-epidemic-related deaths can be neglected in the epidemic timescale and that the incubation period is negligible too. Indicating with letters not in bold the cardinality of each of the compartments, it is taken
\begin{equation}
  S(t_0) + I(t_0) + R(t_0) = N\,,
\end{equation}
where $t_0$ is an initial time, usually with $R(t_0) = 0$. \newline
The model introduces two parameters, $\beta$ and $\gamma$, having dimension of a frequency. Saying $t$ the time variable, $\gamma$ is defined as the fractional removal rate $(1/I)(dR/dt)$ of individuals from the {\em infectious compartment}. Since $SI$ is understood as the number of possible contacts among the infectious and the susceptible individuals, $\beta/N$ is defined as the fractional decrease rate $-(SI)^{-1}(dS/dt)$ of the number of individuals in the {\em susceptible compartment}: it expresses therefore the fractional increment rate of the number of infectious individuals, that is the increment rate of the {\em infectious compartment} \textbf{I}\,, after subtraction of the rate of people entering the {\em removed compartment} \textbf{R}. \newline
Usually one introduces the following non-dimensional variable and new functions:
\begin{equation}
  \alpha \defeq \frac{\beta}{\gamma} \,, \qquad
  x \defeq \gamma t\,,\qquad s(x) \defeq \frac{S(t)}{N}\,, \qquad i(x) \defeq \frac{I(t)}{N}\,,
  \qquad r(x) \defeq \frac{R(t)}{N}\,, \label{redef}
\end{equation}
$\alpha$ called {\em basic reproduction ratio}.
Then the basic equations given by Kermack and McKendrick \cite{Kermack_McKendrick} are written as
\begin{subequations} 
  \begin{align} 
    \frac{ds}{dx}(x) &= -\,\alpha\, i(x)\,s(x) \label{fond.eqs.1 bis}\\
    \frac{di}{dx}(x) &= i(x) (\alpha\, s(x) - 1) \label{fond.eqs.2 bis}\\ 
    \frac{dr}{dx}(x) &= i(x) \label{fond.eqs.3 bis}
  \end{align}
\end{subequations}
with
\begin{equation}
  s(x) + i(x) + r(x) = s(x_0) + i(x_0) + r(x_0) = 1\,, \label{conservation bis}
\end{equation}
and
\begin{equation} 
  s_0 \defeq s(x_0)\,,\qquad i_0 \defeq i(x_0)\,, \qquad r_0 \defeq r(x_0) \equiv 0\,. 
\end{equation}
Using eq.\ref{fond.eqs.3 bis} in eq.\ref{fond.eqs.1 bis} and formally integrating, one gets $s(x) = s_0\,e^{-\alpha\,r(x)}$; using this and eq.\ref{fond.eqs.3 bis} again, from eq. \ref{fond.eqs.2 bis} one easily finds $i(x) = 1 - s_0\,e^{-\alpha\,r(x)} - r(x)$; then from eq.\ref{fond.eqs.3 bis} she/he will obtains
\begin{equation}
  \frac{dr}{dx}(x) = 1 - s_0\,e^{-\alpha\,r(x)} - r(x)\,. \label{the equation}
\end{equation}
This is a transcendental equation, whose solutions one cannot give explicitly in closed analytic form by elementary functions. In their original paper Kermack and McKendrick  themselves (\cite{Kermack_McKendrick}) gave approximate solutions, however without any exhaustive discussion of applicability for various values of the basic reproduction ratio. Quite recently various authors have approached the problem in different ways, but with the same incompleteness (\cite{Ozyapici Bilgeha, Steven Weinstein, Kroeger Schlickeiser, Pakes, Fowler Hollingsworth}).   In the sequel, on the basis of simple, analytic geometry considerations, a novel method is introduced, producing approximate but accurate solutions, given explicitly, piece-wisely, from generalized logistic function (see \cite{Cramer} for a description of the origin of the logistic function and its adoption in bio-assay); due attention is paid for the method to be robust over the whole range of possible known values of $\alpha$, from just above 1 as for influenza, to 1.4-3.9 as for Covid-19, to 3-5 as for SARS, to 5-7 as for polio, to 10-12 as for varicella, to 12-18 as for measles (see for instance \cite{Heesterbeek} and references therein).

\section{Getting the key differential equation} \label{getting eq.}
For the epidemic to spread, the increment rate of the newly infectious individuals must be higher then the increment rate of the newly removed individuals. Dividing eq.\ref{fond.eqs.1 bis} by eq. \ref{fond.eqs.3 bis}\,, one finds that it must be
\begin{equation}
  1 \,<\, -\frac{ds}{dr}(t) \,=\, \alpha\,s(t) \,. \label{sustainability}
\end{equation}
As a matter of fact this condition implies that $i(t)$ increases over time due to eq.\ref{fond.eqs.2 bis}\,.
The functions $s(x)$, $i(x)$ and $r(x)$ are all defined positive and less or equal to 1; consequently it must be $\alpha > 1$ for the epidemic to spread and $s(x)$ turns to be monotonic decreasing according to eq.\ref{fond.eqs.1 bis}\,, while $r(x)$ monotonic increasing according to eq.\ref{fond.eqs.3 bis}. It follows that the function $i(x)$ starts growing due to \ref{sustainability}\,, reaching necessarily a maximum at a time
$t\msubs{M} = x\msubs{M}/\gamma$ such that  
\begin{equation}
 \alpha\, s(x\msubs{M}) =  1  \,,
\end{equation}
then asymptotically decreasing to zero. This implies that the bounded monotonically increasing function $r(x)$ must exhibit a point of inflection at $t\msubs{M}$, after which it bends, increasing slower and slower, finally flattening to some limiting value
\begin{equation}
  r\msubs{\!\infty} \equiv r(+\infty) \le 1\,.  
\end{equation}
So one must have
\begin{equation}
  0 = \lim_{x\to +\infty} \frac{dr}{dx}(x) =
  1 - s_0\,e^{-\alpha\,r\msubs{\!\infty}} - r\msubs{\!\infty}\,, \label{r infty}
\end{equation}
thus getting a transcendental equation for $r\msubs{\!\infty}$. \newline
Conveniently for the following developments, a new function is introduced, namely\sloppy 
\begin{equation} 
  w(x) = 1 - s_0\,e^{-\alpha\,r(x)}\,, \label{w}
\end{equation}
in terms of which eq.\ref{the equation} is re-written as
\begin{subequations} \label{the w eq}
  \begin{align}
    &\frac{dw}{dx} = F[w]\,, \label{the w 1} \\
    &F[w]\,\defeq\, (1-w) \left[\eps + \alpha\,w + \ln(1-w)\right]\,, \label{F[w]} \\
    &\eps\, =\, - \ln(s_0) = - \ln(1-i_0)\,.  \label{the w 2}
   \end{align}
\end{subequations}
Clearly
\begin{equation}
  \bhat{w}\, \defeq\, \lim_{x\to +\infty}w(x) = 1 - s_0\,e^{-\alpha\,r\msubs{\!\infty}}\,
  \,=\, r\msubs{\!\infty} \label{w_hat}
\end{equation}
must be solution of the equation 
\begin{subequations} \label{the w_hat}
  \begin{align} 
    &F[\bhat{w}] = 0\,, \label{boundary condition}
  \end{align}
\end{subequations}
for eq.\ref{r infty} and the fact that
\begin{equation*}
  \frac{dw}{dx} \,=\, s_0\,\alpha\,e^{-\alpha\,r(x)}\,\frac{dr}{dx}\,,
\end{equation*}
so that
\begin{equation*}
  \frac{dw}{dx} \,=\, 0 \quad \Longleftrightarrow \quad \frac{dr}{dx} \,=\, 0\,.
\end{equation*}
The functional $F[w]$ is null in $w=1$, but $\bhat{w}$ cannot be $1$ because $ 0 \le r(x) \le 1 $ and $s_0$ is not null (see eq.\ref{w}\,); thus $\bhat{w}$ must be solution of the equation
\begin{equation}
    \eps + \alpha\,\bhat{w} + \ln(1-\bhat{w}) = 0 \,, \label{to get w_hat}
\end{equation}
which is nothing but eq.\ref{r infty}\,, as can be easily verified. Eq.\ref{to get w_hat} is transcendental and is to be solved numerically; the interval $[0,\,\bhat{w}]$ is the range of $w(x)$ as $x$ runs from $x_0$ to $+\infty$.\newline
The second derivative of $F$, namely
\begin{equation}
  \frac{d^2F}{dw^2}[w] \,=\, -\, 2\, \alpha + \frac{1}{1-w}  
\end{equation}
starts and remains negative from $w=0$, until it reaches the point of inflection $\wflx$, given by
\begin{equation}
  \wflx \,=\, 1 - \frac{1}{2\alpha}\,; \label{wflx}
\end{equation}
then it becomes positive: thus $F[w]$ starts and remains concave until $w=\wflx$; then it becomes convex.
\begin{figure}[ht] 
  \centering
  \includegraphics[width=0.75\textwidth]{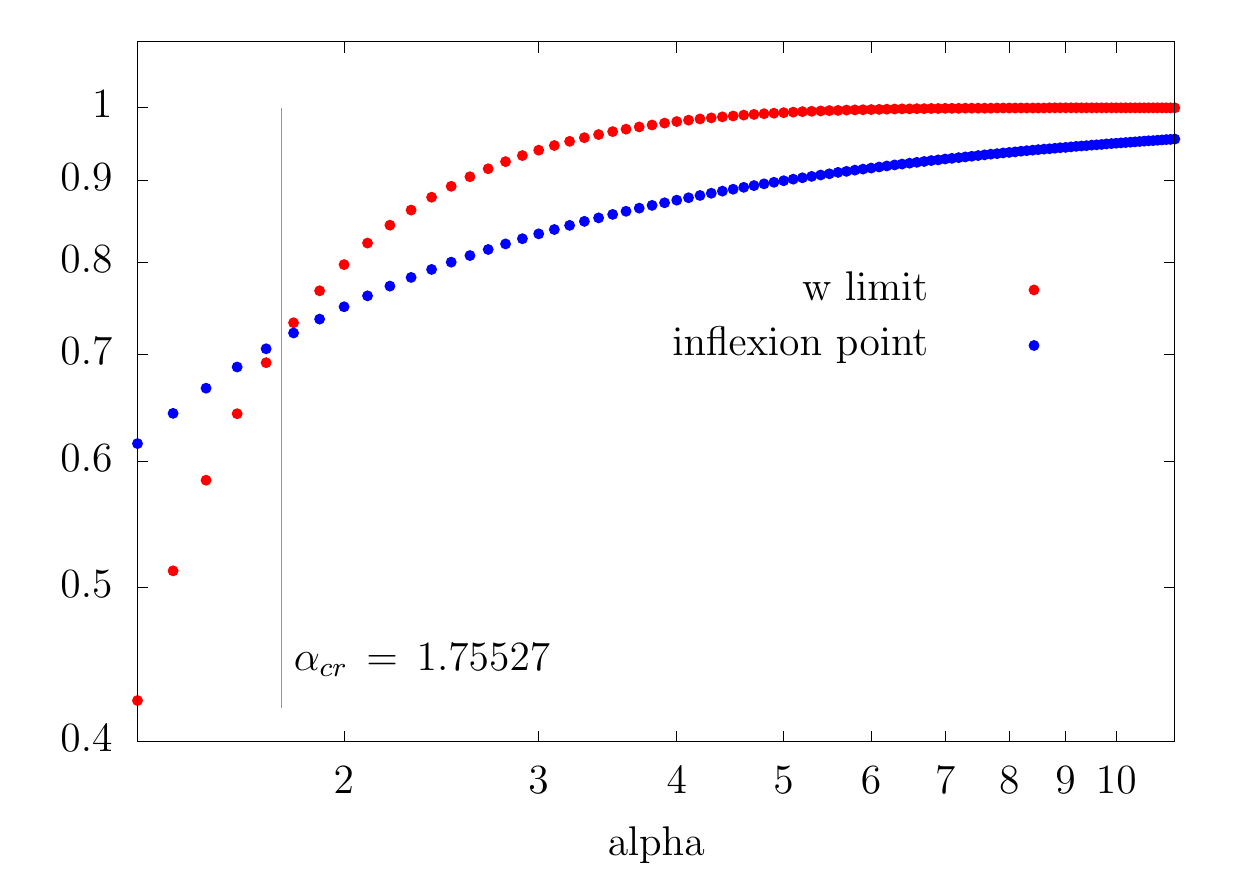}
  \caption{Point of inflection and $\hat{w}$ as a function of $\alpha$.}
  \label{hats and infl}
\end{figure}
Of course, in an interval around its inflection point, $F[w]$ is nearly straight. Fig.\ref{hats and infl} shows how $\bhat{w}$ and $\wflx$ vary as a function of $\alpha$: for $\alpha < \alpha\subs{cr}\simeq 1.75$ one has $\bhat{w} < \wflx$ and consequently $F[w]$ is always concave in the the domain $[0,\,\bhat{w}]$; otherwise it changes from concave to convex after $w=\wflx$. It is worth noting that  as $\alpha$ increases, $\bhat{w}$ (together with $\wflx$) approaches more and more the limiting value 1, namely a region where the log term in $F[w]$ becomes important: this fact is relevant here because such log term, with its argument approaching zero, rises complications in searching for an effective approximation.

\section{Approximating the key differential equation}

\begin{figure}[ht] 
  \centering
  \includegraphics[width=\textwidth]{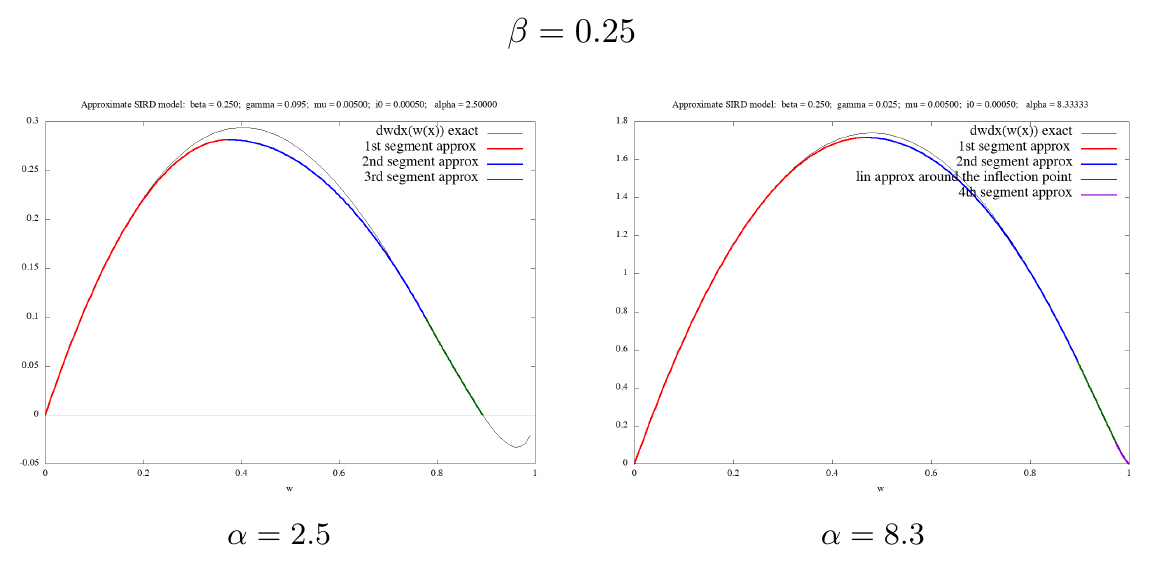} 
  \caption{Examples of two cases, with three approximation stretches on the left (red, blue, green) and four approximation stretches on the right (red, blue, green, magenta)\,.}
  \label{Fig_1.png}
\end{figure}
The idea is to approximate $F[w]$ by few stretches of up to second order po\-ly\-no\-mials, joining continuously each other with the first derivative. Then in each stretch the obtained approximate differential equation becomes analytically and explicitly solvable by a generalized logistic function. For $w \ll 1$\,, it is taken
\begin{equation}
  (1-w)\, \ln(1-w) \approx  -\,w \left( 1 - \frac{1}{2} w\right) \,,
\end{equation}
so that 
\begin{equation}
  \frac{dw}{dx} \approx \eps\,+\,(\alpha-1-\eps)\,w\, -\,
  \left(\alpha - \frac{1}{2}\right)\,w^2 \,\defeq\, \Fsw{1} \,. \label{F1} 
\end{equation}
Fig.\ref{Fig_1.png} shows on the left, in red, this $\Fsw{1}$ segment against $F[w]$ (black curve) for $\alpha = 2.74$ and (consequently) $\bhat{w} \simeq 0.92$\,, extending to its maximum point, which is rather close to the maximum of $F[w]$. Clearly $\Fsw{1}$ is a parabola with axis along the ordinate line, so that the maximum is its vertex.
\newline
Denoting by $\wuu$ and $\wud$ the roots of $\Fsw{1}$, one can write
\begin{subequations} \label{F1bis}
  \begin{align}
    \Fsw{1} \,&=\, -\, A\,(w-\wuu)\,(w-\wud)\,, \label{F1bis1} \\
    A \,&\defeq\, \alpha - \frac{1}{2} \,, \label{F1bis2}
  \end{align}
\end{subequations}
with
\begin{equation}
  \prescript{}{}w^{(1)}_{1/2} \,=\,
  \frac{\alpha-1-\eps\,\pm\sqrt{(\alpha-1-\eps)^2 + 2\,(2\alpha-1)\,\eps}}{2\alpha-1}\,.  \label{W1/2} 
\end{equation}
The vertex is located in
\begin{equation}
  w\msubs{M} \,=\,\frac{\wuu + \wud}{2}\,.  \label{W_M} 
\end{equation}
A new parabola is chosen as the second approximation stretch, tangent to $F[w]$ on its descending side, with axis along the ordinates and the vertex coincident with that of the first segment $\Fsw{1}$: 
\begin{subequations} \label{F2}
  \begin{align}
    &\Fsw{2} \,=\, -\, Z^{\star}\,(w-\wdu)\,(w-\wdd)\,, \label{F2a} \\[5pt]
    &\frac{\wuu + \wud}{2} \,=\, w\msubs{M} \,=\,\frac{\wdu + \wdd}{2}\,, \label{F2b}\\[5pt]
    & -\, A\,(w\msubs{M}-\wuu)\,(w\msubs{M}-\wud) \,=\,
    -\, Z^{\star}\,(w\msubs{M}-\wdu)\,(w\msubs{M}-\wdd)\,, \label{F2c} \\[3pt]
    &\Fsw{2} \,=\, F[w]\,, \label{F2d} \\[3pt]
    &\frac{\delta F\!\supt{(2)}}{\delta w\quad}[w(x)] \,=\, \frac{\delta F}{\delta w}[w(x)] \,. \label{F2e}
  \end{align}
\end{subequations}
Equations \ref{F2b} and \ref{F2c} impose that the two stretches have in common their vertexes, located in $w=w\msubs{M}$; the system of the last two equations states the conditions for $\Fsw{2}$ to be tangent to $F[w]$. It is convenient expressing $Z^{\star}$, appearing in eq.\ref{F2c}\,, in terms of the unknown tangency point $w^{\star}$ using eq.\ref{F2e}\,, so that consequently one solves eq.\ref{F2d} for $w^{\star}$.\newline
Namely, introducing
\begin{subequations} \label{a}
  \begin{align}
    &\dewu \,\defeq\, \frac{\wuu - \wud}{2} \,, \label{dw1}\\
    &\dewd \,\defeq\, \frac{\wdu - \wdd}{2} \,, \label{dw2}
  \end{align}
\end{subequations}
due to eq.\ref{F2c} one can write
\begin{equation} \label{A-Z}
  Z^{\star}\,(\dewd)^2 \,=\, A\,\,(\dewu)^2   \,,  
\end{equation}
while from eq.\ref{F2e} and eq.\ref{F2d} one has
\begin{subequations} \label{Z-wstar}
  \begin{align}
    &(1-w^{\star}) \left[\eps + \alpha\,w^{\star} + \ln(1-w)\right] \,=\,
    -\,Z^{\star}\,(w-w\msubs{M})^2\,+\,A\,(\dewu)^2\,, \label{Z-wstar 1}
    \\[7pt]
    &Z^{\star} \,=\,
    \frac{1+\eps+2\alpha w^{\star}+\ln{(1-w^{\star})}-\alpha}{2\,(w^{\star}-w\msubs{M})}\,.
    \label{Z-wstar 2}
  \end{align}
\end{subequations}
Using this expression for $Z^{\star}$ in eq.\ref{Z-wstar 1}\,, one obtains a transcendental ordinary equation for $w^{\star}$\,, to be solved numerically:
\begin{align} \label{the w_star}
  2 \eps \,+\, (\alpha -\eps -1)\,\!w\msubs{M} \,-\, 2A(\dewu)^2 \,&+\,
  (\alpha-\eps-2 \alpha w\msubs{M}+1)\,w^{\star} \\ 
  \,&+\, (2-w^{\star}-w\msubs{M})\,\ln(1-w^{\star}) \,=\,0\,.\nonumber
\end{align}
Using $w^{\star}$ so obtained, one gets $Z^{\star}$ from eq.\ref{Z-wstar 2} and finally $\wdu$ and $\wdd$ via eq.\ref{A-Z} and eq.\ref{F2b}\,. In fig.\ref{Fig_1.png}, on the left, the second segment for $\alpha = 2.6$ is shown in blue, extending from $w\msubs{M}$ to the point of tangency of the successive approximation segment still to be chosen. 
\begin{figure}[ht] 
  \centering
  \includegraphics[width=\textwidth]{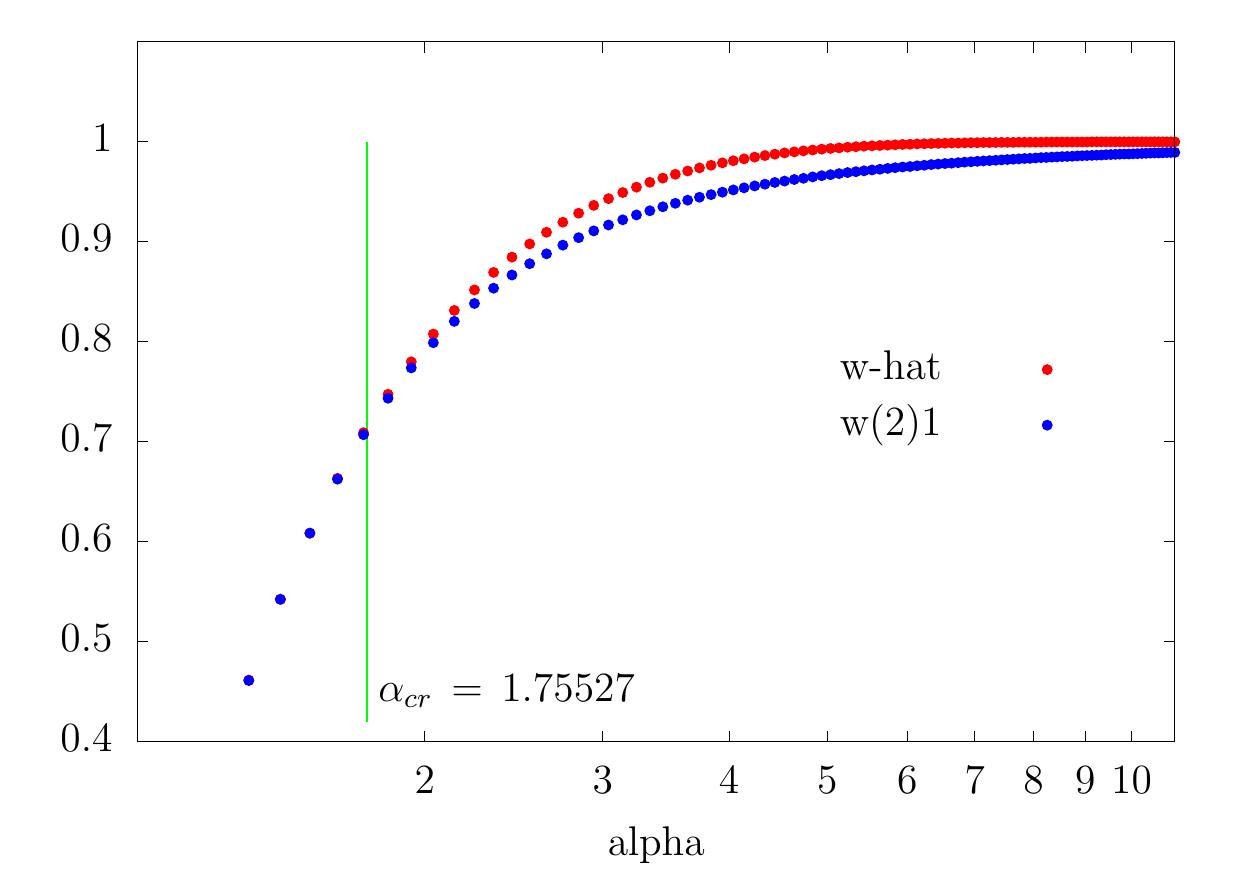}
  \caption{$\wdu$ and $\hat{w}$ as functions of $\alpha$.}
  \label{Xi1s}
\end{figure}
\newline
With reference to the discussion before the end of Section \ref{getting eq.}\,, it should be noted that $F[w]$ remains concave up to $w = \bhat{w}$ when $\alpha \le \alpha\subs{cr}$\,, while it happens that the root $\wdu$ of $\Fsw{2}$ (see fig.\ref{Xi1s}\,) remains very close to $\bhat{w}$\,: this suggests in that range of $\alpha$ values replacing the above $\Fsw{2}$ by a different arc of parabola $\fsw{2}$\,, keeping its vertex in common with $\Fsw{1}$ as $\Fsw{2}$ does, but just ending in $\bhat{w}$, thus imposing the constraint $\wdu = \bhat{w}$ instead of the tangency to $F[w]$. \newline
Then for $\alpha \le \alpha\subs{cr}$
\begin{subequations} \label{f2}
  \begin{align}
    &\fsw{2} \,=\, -\, Z\,(w-\wfu)\,(w-\wfd)\,, \label{f2a} \\[5pt]
    &\frac{\wuu + \wud}{2} \,=\, w\msubs{M} \,=\,\frac{\wfu + \wfd}{2}\,, \label{f2b}\\[5pt]
    & -\, A\,(w\msubs{M}-\wuu)\,(w\msubs{M}-\wud) \,=\,
    -\, Z\,(w\msubs{M}-\wfu)\,(w\msubs{M}-\wfd)\,, \label{f2c} \\[3pt]
    &Z = \left(\alpha - \frac{1}{2}\right)\,\frac{\big(\dewu\big)^2}{(\bhat{w}-w\msubs{M})^2}\,, \label{f2d} \\[3pt]
    &\wfu \,=\, \bhat{w}\,, \quad \wfd \,=\, 2 w\msubs{M}-\bhat{w}\,, \quad \dewf = \bhat{w}- w\msubs{M}\,.\label{f2e}
  \end{align}
\end{subequations}
\par
For $\alpha\subs{cr} < \alpha \le 6$\, $F[w]$  is almost always concave, ending roughly as a straight line when approaching $\bhat{w}$. In this range of $\alpha$'s one keeps $\Fsw{2}$\,, but completes the approximation through a new parabola, requiring it to be tangent to $\Fsw{2}$ and to reach $\bhat{w}$ along the tangent to $F[w]$ in $\bhat{w}$; an alternative is the ray originating in $\bhat{w}$, tangent to $\Fsw{2}$\,. The latter is settled by
\begin{subequations}
  \begin{align}
    &L[w] \,\defeq\, -2\,u\,Z^{\star}\,(w-\bhat{w})\,, \label{linear} \\[5pt]
    \mathlarger{\backepsilon}\,'\quad 
    \Bigg\{\; L[w] \,-\,& \Fsw{2} \,=\,0 \quad\wedge\quad
    \Delta\left(L[w]-\Fsw{2}\right) = 0 \;\Bigg\}\,,
  \end{align}
\end{subequations}
where $\Delta\left(L[w]-\Fsw{2}\right)$ is the discriminant of the second order algebraic equation
$L[w] \,-\, \Fsw{2} \,=\,0 $\,, set to zero to assure $L[w]$ to be tangent to $\Fsw{2}$\,. The appropriate solution for $u$ is
\begin{equation}
  u_- \,=\, \bhat{w} - w\msubs{M} -
  \sqrt{(\bhat{w} - w\msubs{M})^2 - (\dewd)^2}  \,.  \label{u-} 
\end{equation}
The problem with this approximation is that, looking for instance at the function $r(x)$ obtained from $w(x)$, it gets unacceptably overestimated in the region where it bends to reach the asymptotic value as $x\rightarrow +\infty$: this is because $L[w]$ necessarily remains below $F[w]$ due to the concavity of the latter.  \newline
The quadratic alternative is defined by
\begin{subequations}
  \begin{align}
    &\Fsw{3} \,\defeq\, -2\,\lambda\,(w-\bhat{w})\,+\,\sigma\,(w-\bhat{w})^2\,, \label{F3}\\
    &\,\lambda = \left.(\,\Fsw{2}\,)'\right\vert_{w= \bhat{w}} \;=\;
    \frac{1-\alpha\,(1-\bhat{w})}{2}\,, \\
    \mathlarger{\backepsilon}\,'\quad
    & \Bigg\{\;\, \Fsw{3} \,-\, \Fsw{2} \,=\,0 \quad\wedge\quad
    \Delta\left(\,\Fsw{3}-\Fsw{2}\right) = 0 \;\,\Bigg\}\,, \label{tangF3}
  \end{align}
\end{subequations}
where ``prime'' stands for derivative and $\Delta\left(\,\Fsw{3}-\Fsw{2}\right)$ is the discriminant of the second order algebraic equation $\Fsw{3} \,-\, \Fsw{2} \,=\,0$\,, set to zero so to assure $\Fsw{3}$ to be tangent to $\Fsw{2}$\,. In this case, however, with respect to using  $L[w]$\,, one has the opposite effect on $r(x)$, because the given choice for $\lambda$ forces $\Fsw{3}$ to stay somewhat above $F[w]$. \newline
The solution is to keep the quadratic alternative, but replacing the previous value of $\lambda$ by a compromise one, defined through
\begin{equation}
  \ecirc{\lambda} \,\defeq\,  \tan\Bigl(\arctan(-2\,\lambda)\Bigr) \,+\,
  \tan\left(\frac{\arctan(-2\,\lambda)-\arctan(-\,2\,u_-\,Z^{\star})}{2}\right)\,. \label{lambdao}
\end{equation}
Then the parameter $\sigma$ in \ref{F3} is set by means of the the condition \ref{tangF3}\,:
\begin{subequations} \label{F3 solution}
  \begin{align}
    &\qquad\qquad\quad\sigma \,=\,
    \frac{Z^{\star}\,h \,-\,g^2}{2\,\bhat{w}\,g \,-\,h\,-\,Z^{\star}\,{\bhat{w}}^2}\,, \label{sigma}\\
    &\,g \,=\, Z^{\star}\,w\msubs{M} \,+\, \ecirc{\lambda}\,,\qquad
    \,h \,=\, Z^{\star} \wdu\wdd\,+\,2\,\ecirc{\lambda}\,\bhat{w}\,, \label{hg} \\
    & \qquad\qquad\qquad\, \ecirc{w} \,=\,
    \frac{\sigma\,\bhat{w} \,+\, g}{\sigma\,+\,Z^{\star}}\,, \label{wcirc}
  \end{align}
\end{subequations}
where $\ecirc{w}$ is the tangency point of $\Fsw{3}$ to $\Fsw{2}$.\newline
So, for $\alpha\subs{cr} < \alpha <= 6$ the third and last approximation segment is given by \ref{F3}\,, with $\lambda$ replaced by $\ecirc{\lambda}$, extending from $\ecirc{w}$ to $\bhat{w}$. \newline 
For $w > 6$ the convexity trait of $F[w]$, following the almost straight stretch around $w\subs{flx}$\,, gets more and more included in the domain $[0,\,\bhat{w}]$\,, because $\bhat{w}$ increases with $\alpha$. Then, the solution adopted is to introduce a linear segment $T[w]$ parallel to the tangent in $\wflx$ to $F[w]$ and tangent to $\Fsw{2}$ in a point that will be denoted  $\tilde{w}$; this linear segment will be continued by a new parabola $\Fsw{4}$, which is similar to $\Fsw{3}$, thus ending in $\bhat{w}$, but tangent to $T[w]$.  Namely
\begin{subequations}\label{the T}
  \begin{align}
    T[w] \,&\defeq\, - 2\,\tilde{f}\,w \,+\,\tilde{I}\,, \label{Tlin}\\
    -2 \tilde{f} \,&\defeq\, \left.F'[w]\,\right\vert_{w=\wflx} \,=\,
    \ln(2 \alpha) \,-\, \alpha \,-\, \eps\,, \label{ftilde}\\
    \mathlarger{\backepsilon}\,'\quad 
    \Bigg\{\; T[w] \,&-\, \Fsw{2} \,=\,0 \quad\wedge\quad
    \Delta\left(T[w]-\Fsw{2}\right) = 0 \;\Bigg\}\,,
  \end{align}
\end{subequations}
giving
\begin{subequations}\label{the tilde}
  \begin{align}
    &\tilde{I} \,=\, Z^{\star}\,
    \left[ {\tilde{w}}^2 \,-\, w\msubs{M}^2
      \,+\, (\dewd)^2 \right]\label{Itilde} \\
    &\qquad\qquad \tilde{w} \,=\, w\msubs{M} \,+\,\frac{\tilde{f}}{Z^{\star}} \,. \label{wtilde}
  \end{align}
\end{subequations}
Then the $\Fsw{4}$ approximation stretch, constrained to end in $\bhat{w}$ and to be tangent to $T[w]$ in a point $\ecup{w}$ chosen by trial and error optimization, is given by:
\begin{subequations}
  \begin{align}
    &\Fsw{4} \,\defeq\,
    -2\,\ecup{\lambda}\,(w-\bhat{w})\,+\,\ecup{\sigma}\,(w-\bhat{w})^2\,, \label{F4}\\
    & \ecup{w} \,\defeq\, (1-z)\,\wflx \,+\, z\,\bhat{w}\,, \qquad
    z = 0.575\,, \label{wu}\\
    \mathlarger{\backepsilon}\,'\quad 
    \Bigg\{&\;\, \Fsw{4} \,-\, T[w] \,=\,0 \quad\wedge\quad
    \Delta\left(\,\Fsw{4}-T[w]\right) = 0 \;\,\Bigg\}\,,
  \end{align}
\end{subequations}
giving
\begin{subequations} \label{the u}
  \begin{align}
    &\ecup{\lambda} \,=\,
    \tilde{f}\,+\,\frac{2\bhat{w}\,\!\tilde{f}\,-\,\tilde{I}}{\ecup{w}\,-\,\bhat{w}}\,,
    \label{lambdau}\\
    &\ecup{\sigma} \,=\, \frac{2 \bhat{w} \tilde{f}\,-\,\tilde{I}}{(\ecup{w}\,-\,\bhat{w})^2}\,.
    \label{sigmau}
  \end{align}
\end{subequations}

\section{The approximate analytic solution}
For each of the above approximation segments a differential equations is defined of the type
\begin{equation}
 \frac{dw}{dx}(x) \,=\, \mathcal{F}[w(x)]\,,  \label{the_eq.}
\end{equation}
where $\mathcal{F}[w]$ is one of $\Fsw{i}$\, $(i=1,2,3,4)$ or $\fsw{2}$ or $T[w]$, with given $\alpha$ and $\beta$ parameters (or $\beta$ and $\gamma$) and initial conditions. For $\mathcal{F}[w] = \Fsw{1}$\,, from the definition in eq.\ref{w}\,, the initial condition is $w(x\msubs{0}) = 1-s\msubs{0} = i\msubs{0}$\, ($x\msubs{0} = 0$ without loss of generality), while for each of the remaining approximation segments it is given by the value of the respective preceding segment at the junction point. Since $\mathcal{F}\![w]$ is at most a second order polynomial, eq.\ref{the_eq.} is indeed quite trivially solved, giving a generalized logistic function.
\newline
\newline
For $\mathcal{F}\![w] = \Fsw{1}$:
\begin{subequations} \label{sol_w1} 
  \begin{align}
    &\qquad\quad\;\; \wU(x)\,=\,\frac{ \wuu\,+\,\wud\,k\;\nexp{x}{\gamma \tau\msubs{1}} }
    { 1\,+\,k\;\nexp{x}{\gamma \tau\msubs{1}} }\,, \\
    &k\,=\, \frac{\wuu\,-\,i\msubs{0}}{i\msubs{0}\,-\,\wud}\,, \qquad
    \tau\msubs{1} \,=\,
    \frac{1}{\gamma\cdot (\alpha\,-\,\nicefrac{1}{2})\, (\wuu\,-\,\wud)} \,. 
  \end{align}
\end{subequations}
\vspace{5pt}\par
For $\mathcal{F}\![w] = \fsw{2}$\,, thus $\alpha \le \alpha\subs{cr}$:
\begin{subequations} \label{sol_wf} 
  \begin{align}
    &\qquad\qquad \wf(x)\,=\,\frac{ \bhat{w}\,+\,(2 w\msubs{M}- \bhat{w}) \;\nexp{(x-x\msubs{M})}{\gamma \tau\msubs{f}} }
    { 1\,+\,\nexp{(x-x\msubs{M})}{\gamma \tau\msubs{f}} }\,, \\
    &x\msubs{M}\,=\,\gamma\,\tau\msubs{1}\ln{(k)}\;\, \suchthat\;\,
    \wU(x\msubs{M})\,=\,w\msubs{M}\,,\quad
    \tau\msubs{f} \,=\, \frac{\dewf}{\dewu} \, \tau\msubs{1}
    \;>\; \tau\msubs{1}\,. 
  \end{align}
\end{subequations}
\vspace{5pt}\par
For $\mathcal{F}\![w] = \Fsw{2}$\,, thus $\alpha > \alpha\subs{cr}$:
\begin{subequations} \label{sol_w2} 
  \begin{align}
    &\qquad\qquad \wD(x)\,=\,\frac{ \wdu\,+\,\wdd\;\nexp{(x-x\msubs{M})}{\gamma \tau\msubs{2}} }
    { 1\,+\,\nexp{(x-x\msubs{M})}{\gamma \tau\msubs{2}} }\,, \\
    &x\msubs{M}\,=\,\gamma\,\tau\msubs{1}\, \ln{(k)}\; \suchthat\;
    \wU(x\msubs{M})\,=\,w\msubs{M}\,,\quad
    \tau\msubs{2} \,=\, \frac{\dewd}{\dewu} \, \tau\msubs{1}
    \;>\; \tau\msubs{1}\,. 
  \end{align}
\end{subequations}
\vspace{5pt}\par
For $\mathcal{F}\![w] = \Fsw{3}$\,, thus $\alpha\subs{cr} < \alpha \le 6$ :
\begin{subequations} \label{sol_w3} 
  \begin{align}
    &\qquad \wTre(x)\,=\,
    \frac{ \bhat{w}\,-\,(\bhat{w}\,+\,{2\ecirc{\lambda}}/{\sigma})\;\ecirc{\phi}\;
      \nexp{(x-\ecirc{x})}{\gamma \tau\msubs{3}} } 
    { 1\,-\,\ecirc{\phi}\,\nexp{(x-\ecirc{x})}{\gamma \tau\msubs{3}} }\,, \\
    &\ecirc{x} \,=\, \gamma\,x\msubs{M} \,+\, \gamma\,\tau\msubs{2}\,
    \ln{\left(\frac{\ecirc{w}-\wdd}{\wdu-\ecirc{w}}\right)}\quad
    \suchthat\; \wD(\ecirc{x}) = \ecirc{w}\,,\\
    &\qquad\quad \ecirc{\phi}\,=\,
    \frac{\bhat{w}\,-\,\ecirc{w}}{\bhat{w}\,-\,
      \ecirc{w}\,+\,\frac{2\ecirc{\lambda}}{\sigma}}\;,
    \qquad \tau\msubs{3} \,=\, \frac{1}{2\,\ecirc{\lambda}\,\gamma}\,. 
  \end{align}
\end{subequations}
\vspace{5pt}\par
For $\mathcal{F}\![w] = T[w]$ \,, thus $\alpha > 6$ (see \ref{the T} and \ref{the tilde}): 
\begin{subequations} \label{Linw} 
  \begin{align}
    &\qquad\qquad \wT(x)\,=\,\frac{1}{2\,\tilde{f}} \,
    \left[ \tilde{I} \,-\, (\tilde{I}\,-\,2\,\tilde{w}\,\tilde{f})
      \;\nexp{(x-\tilde{x})}{\gamma \tilde{\tau}} \right]   \,, \\
    &\tilde{\tau} \,=\, \frac{1}{2\,\tilde{f}\,\gamma}\,, 
      \qquad \tilde{x} \,=\, \gamma\,x\msubs{M} \,+\,
      \gamma\,\tau\msubs{2}\,
      \ln{\left(\frac{\tilde{w}-\wdd}{\wdu-\tilde{w}}\right)}\;
      \suchthat\; \wD(\tilde{x}) = \tilde{w}\,. 
  \end{align}
\end{subequations}
\vspace{5pt}\par
Finally for $\mathcal{F}\![w] = \Fsw{4}$  thus $\alpha > 6$\,:
\begin{subequations} \label{sol_w4} 
  \begin{align}
    &\; \wQ(x)\,=\,
    \frac{ \bhat{w}\,-\,(\bhat{w}\,+\,{2\,\ecup{\lambda}}/{\ecup{\sigma}})\;\ecup{\phi}\;
      \nexp{(x-\ecup{x})}{\gamma \tau\msubs{4}} } 
         { 1\,-\,\ecup{\phi}\,\nexp{(x-\ecup{x})}{\gamma \tau\msubs{4}} }\,, \\
    &\ecup{x} \,=\,\gamma\,\tilde{x}\,+\,
         \gamma\,\tilde{\tau}\,\ln{\left(\frac{\tilde{I}\,-\,2\,\tilde{f}\,\tilde{w}}
           {\tilde{I}\,-\,2\,\tilde{f}\,\ecup{w}}\right)} \quad
         \suchthat\; \wT(\ecup{x}) = \ecup{w}\\
         &\quad\; \ecup{\phi}\,=\,
         \frac{\bhat{w}\,-\,\ecup{w}}{\bhat{w}\,-\,\ecup{w}\,
           +\,\frac{2\ecup{\lambda}}{\ecup{\sigma}}}\;,
    \qquad \tau\msubs{4} \,=\, \frac{1}{2\,\ecup{\lambda}\,\gamma}\,. 
  \end{align}
\end{subequations}
It is convenient to introduce
\begin{equation}
  \bcheck{r}(t) \,\defeq\, r(\gamma\,t)\,, \quad
  \bcheck{i}(t) \,\defeq\, i(\gamma\,t)\,, \quad
  \bcheck{s}(t) \,\defeq\, s(\gamma\,t)\,, \quad
  \bcheck{w}(t) \,\defeq\, w(\gamma\,t)\,,\quad \text{etc.}\,,\label{bcheck}\,.
\end{equation}
Then, from eq.\ref{w} one has
\begin{equation} \label{r(t)}
  \bcheck{r}(t) \,=\, \frac{1}{\alpha}\,\ln\frac{1\,-\,i\msub{0}}{1-w(\gamma\,t)}\,,
\end{equation}
so that
\begin{equation*} 
  \bcheck{i}(t)\,=\,\frac{d\bcheck{r}}{dt}(t) \,=\,
  \frac{1}{\alpha}\, \left[\frac{1}{1-w(x)}\,\frac{dw}{dx}(x)\right]_{x=\gamma t}\,. 
\end{equation*} 
On the other hand eq.\ref{the w eq} implies
\begin{equation*}
  \frac{1}{1-w}\,\frac{dw}{dx} \,=\, \alpha\,w \,-\, \ln\frac{1\,-\,i\msub{0}}{1-w}
\end{equation*}
and consequently (see eq.\ref{r(t)})
\begin{equation} \label{i(t)}
  \bcheck{i}(t)\,=\,
  \left[w(x)\,-\,\frac{1}{\alpha}\,\ln\frac{1\,-\,i\msub{0}}{1-w(x)}\right]_{x=\gamma t}
  \,=\, \bcheck{w}(t) - \bcheck{r}(t)\,.
\end{equation}
Finally, of course, due to \ref{conservation bis},:
\begin{equation} \label{s(t)}
  \bcheck{s}(t) \,=\, 1\,-\, \bcheck{i}(t)\,-\, \bcheck{r}(t)\,=\, 1 - \bcheck{w}(t)\,.
\end{equation}
\begin{figure}[ht] 
  \centering
  \includegraphics[width=\textwidth]{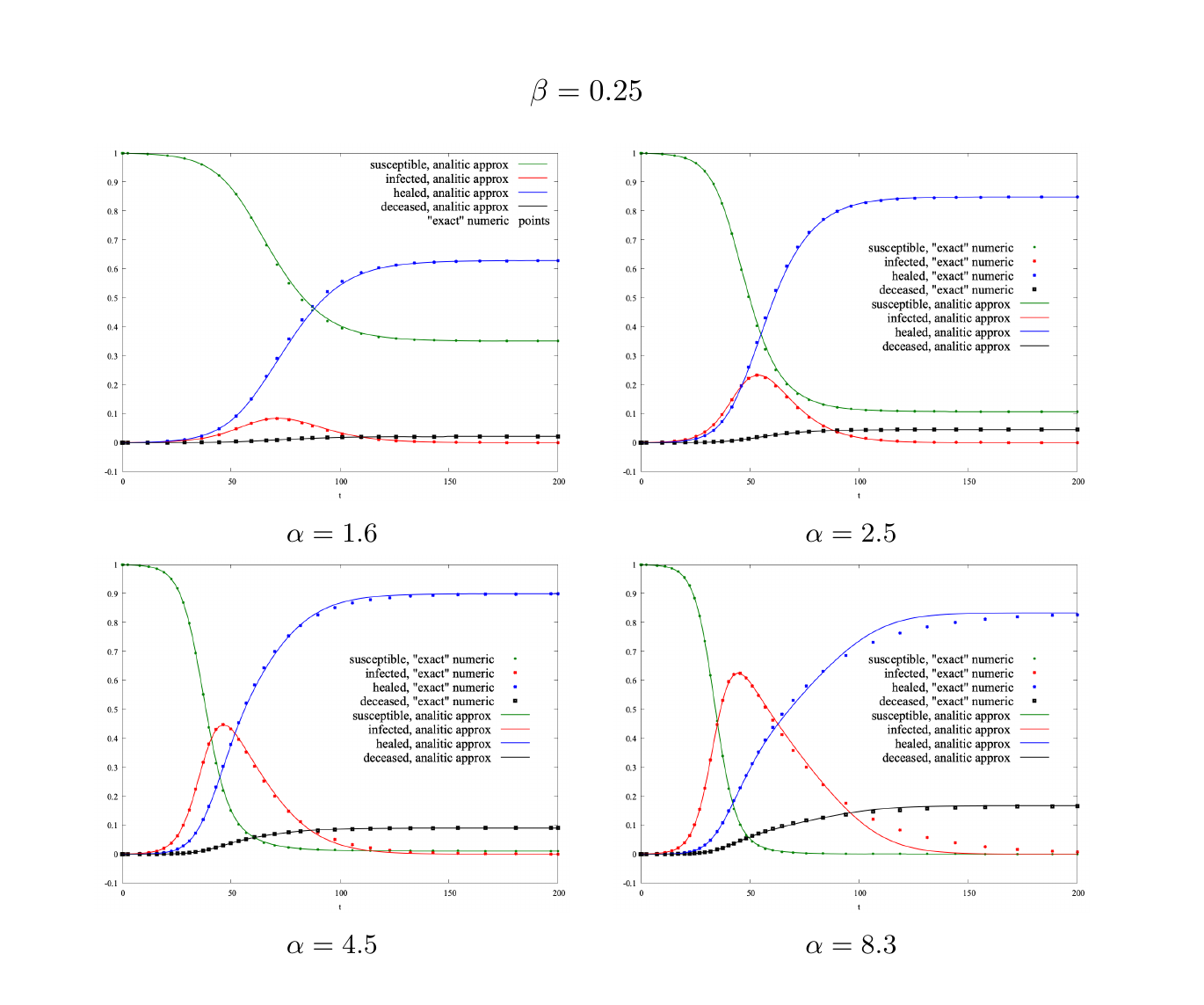}
  \caption{Comparison of ``exact'' numerical solutions and approximate solutions for the SIRD model.}
  \label{Fig: alphas}
\end{figure}
\vspace{5pt}\par
In the case of the SIRD model one defines
\begin{subequations} \label{r=h+d}
  \begin{align}
    &\qquad\qquad\qquad\qquad\quad \bcheck{r} \,=\, \bcheck{h} \,+\, \bcheck{d}\,, \\
    &\gamma \rightarrow \gamma \,+\, \mu\quad\; \text{so that}\quad\;
    \bcheck{h} = \frac{\gamma}{\gamma+\mu}\, \bcheck{r} \quad \text{and} \quad
    \bcheck{d} = \frac{\mu}{\gamma+\mu}\, \bcheck{r} \,.
  \end{align}
\end{subequations}
Fig.\ref{Fig: alphas} shows a comparison between the numerical ``exact'' solutions of the SIRD model and the approximate solutions of this work with $\beta = 0.25$ and $\alpha = 1.6\,,\,2.5\,,\,4.5\,,\,8.3$\,.
\par
Imitating a formal expression typical of computing languages\footnote{$(a \le b)$\,? then $c=f$\,: otherwise\, $c=g$}, the result for $w$ can be summarized as follows:
\vspace{-5pt}
\begin{subequations} \label{w(t)}
  \begin{align}
    &\text{for}\; \alpha \le \alpha\subs{cr} \\
    &\qquad \bcheck{w}(t) \,=\, (t \le t\msubs{M})\,? \bwU(t)\,:\,\bwf(t)
    \label{a_cr <= a} \\[5pt]
    &\text{for}\; \alpha\subs{cr} < \alpha \,\le\, 6\,: \nonumber \\ 
    &\qquad \bcheck{w}(t) \,=\, (t \le t\msubs{M})\,? \bwU(t)\,:\,
    \Big(\,(t \le \ecirc{t})\,? \bwD(t)\,:\, \bwT(t) \,\Big) \label{a<=6} \\[5pt]
    &\text{for}\; \alpha \,>\, 6\,: \nonumber \\
    &\qquad \bcheck{w}(t) \,=\, (t \le t\msubs{M})\,? \bwU(t)\,:\, \nonumber \\
    &\qquad\qquad\qquad\qquad \bigg(\,(t \le \tilde{t}\,)\,?\,\bwD(t)\,:\,
    \Big(\,(t\le\ecup{t})\,?\,\bwL(t)\,:\,\bwQ(t)\,\Big)\,\bigg)\,.
  \end{align}
\end{subequations}
Similarly for $\bcheck{s}(t)$\,, $\bcheck{i}(t)$\,, $\bcheck{h}(t)$ and $\bcheck{d}(t)$\,.
\\[10pt]
{\bf In practice} one does:
\vspace{-3pt}  
\begin{itemize}
\item solve numerically the transcendental ordinary eq.\ref{to get w_hat} to get $\bhat{w}$;
\item use eq.\ref{W1/2} and eq.s\ref{sol_w1} to get $\wU(x)$ as in eq.\ref{sol_w1};
\item for $\alpha \le \alpha\subs{cr}$ use  eq.\ref{W_M}\,, eq.\ref{f2d} and \ref{f2e} to get $\wf(x)$ as in eq.\ref{sol_wf}\,;
\item for $\alpha > \alpha\subs{cr}$ use eq.\ref{the w_star}\,, eq.\ref{Z-wstar 2}\,, eq.\ref{W_M}\,, eq.\ref{dw2}\,, eq.\ref{A-Z} and eq.s\ref{sol_w2} to get $\wD(x)$ as in eq.\ref{sol_w2}\,;
\item for $ \alpha\subs{cr} < \alpha <= 6$ use eq.\ref{lambdao}\,, eq.\ref{sigma} and eq.\ref{hg}\,, eq.\ref{wcirc} and finally eq.s\ref{sol_w3} to get $\wTre(x)$ as in eq.\ref{sol_w3}\,;
\item for $\alpha > 6$ use eq.\ref{ftilde}\,, eq.s\ref{the tilde} and eq.s\ref{Linw} to get $\wT(x)$ as in eq.\ref{Linw}\,;
\item for $\alpha > 6$ use eq.\ref{wflx}\,,  eq.s\ref{wu}\,, eq.s\ref{the u} and eq.s\ref{sol_w4} to get $\wQ(x)$ as in eq.\ref{sol_w4}\,;
\item eventually use eq.\ref{r(t)}\,, eq.\ref{i(t)}\,, eq.\ref{s(t)}\,, eq.\ref{r=h+d}\,.
\end{itemize}
The four plots in fig.\ref{Fig: alphas} are produced by a C++ code implementing the above steps, then sending the produced analytic function to the graphing utility ``gnuplot'': the C++ code could be re-used easily to fit-study data.

\section{A useful feature}
The equation of the first approximation segment can be re-written as
\begin{figure}[ht]
  \centering
  \includegraphics[width=1.0\textwidth]{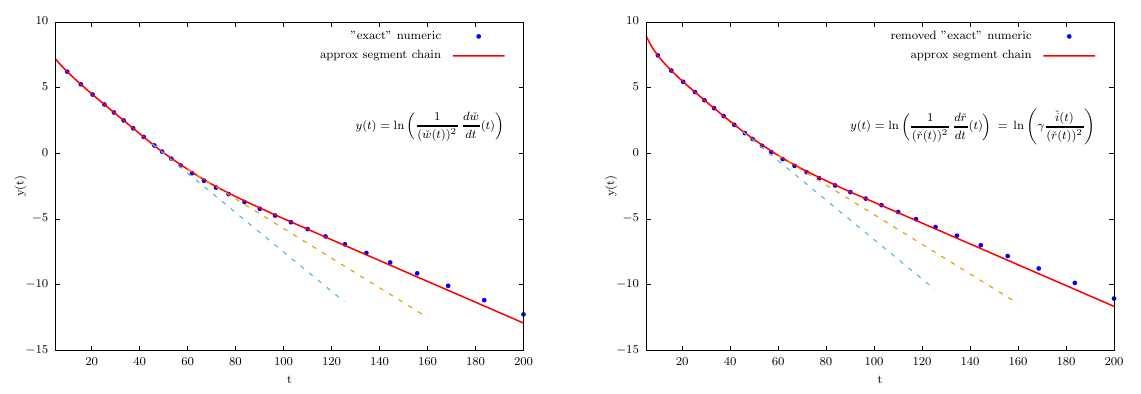} 
  \caption{}
  \label{Fig:log_theory}
\end{figure}
\begin{equation}
  \frac{1}{{\wU}^2}\frac{d\!\wU}{dx\;\;}\,=\, -\, \frac{A}{{\wU}^2}\,(\wU-\wud -\dewu)\,(\wU-\wud)\,.
\end{equation}
Using the explicit solution eq.\ref{sol_w1}\,, one has
\begin{equation}
  \wU-\wud \,=\, 
  \frac{2\,\dewu}{ {\left[1\,+\,k\;\nexp{(x-x\msubs{0})}{\gamma \tau\msubs{1}}\right]}^2 }\,.
\end{equation}
and consequently
\begin{equation}
  \frac{1}{{\wU}^2}\frac{d\!\wU}{dx\;\;}\,=\, 4\,A\,k\, \frac{(\dewu)^2}{{\wuu}^2}\,
  \frac{\nexp{(x-x\msubs{0})}{\gamma \tau\msubs{1}}}{ {\left[1\,+\,\frac{\wuu}{\wud}\,k\;\nexp{(x-x\msubs{0})}{\gamma \tau\msubs{1}}\right]}^2 }\,.
\end{equation}
Typically
\begin{equation}
  \left|\frac{\wuu}{\wud}\right|\, \ll\, 1\quad \text{and}\quad \left|\frac{\wuu}{\wud}\,k\right|\, \lesssim\, 1\,,
\end{equation}
but anyway with $t-t\msubs{0}$ greater then some $\tau\msubs{1}$'s, in the end one can write 
\begin{equation}
  \ln\left(\frac{1}{({\wUc})^2}\frac{d\wUc}{dt}\right)(t)\,\simeq\,
  \ln(4\,A\,\gamma\,k)\;-\,\frac{t-t\msubs{0}}{\tau\msubs{1}}\,.
\end{equation}
Analogous results hold for all the approximation stretches in the different $\alpha$ intervals as summarized in eq.s\ref{w(t)}\,; for instance, with $t-\ecirc{t}$ greater enough then $\tau\msubs{3}$, one has
\begin{equation}
  \ln\left(\frac{1}{({\wTc})^2}\frac{d\wTc}{dt}\right)(t)\,\simeq\,    \ln\left[4\,\sigma\,\gamma\,\ecirc{\phi}\left(\frac{2\ecirc{\lambda}}{\sigma\bhat{w}}\right)^2\right]\;
    -\,\frac{t-\ecirc{t}}{\tau\msubs{3}}\,.
\end{equation}
These piecewise linear behaviors can be seen in fig.\ref{Fig:log_theory} for
$\alpha = 2.6$\,. The plot on the left shows the numerical solution of the exact equation, compared with the corresponding approximate analytic solution: it is worth recalling (see eq.\ref{i(t)}\,) that $w(x)\,=\,r(x)+i(x)$, so that $w$ is directly related to the data. The plot on the right shows that the function $\bcheck{r}(t)$ of the {\it removed} individuals exhibits an analogous behavior: since in the SIRD model the $\bcheck{d}(t)$ function is a fraction of $\bcheck{r}(t)$, then one has the analogous behavior for the function of the deceased individuals.
\begin{figure}[ht]
  \centering
  \includegraphics[width=0.80\textwidth]{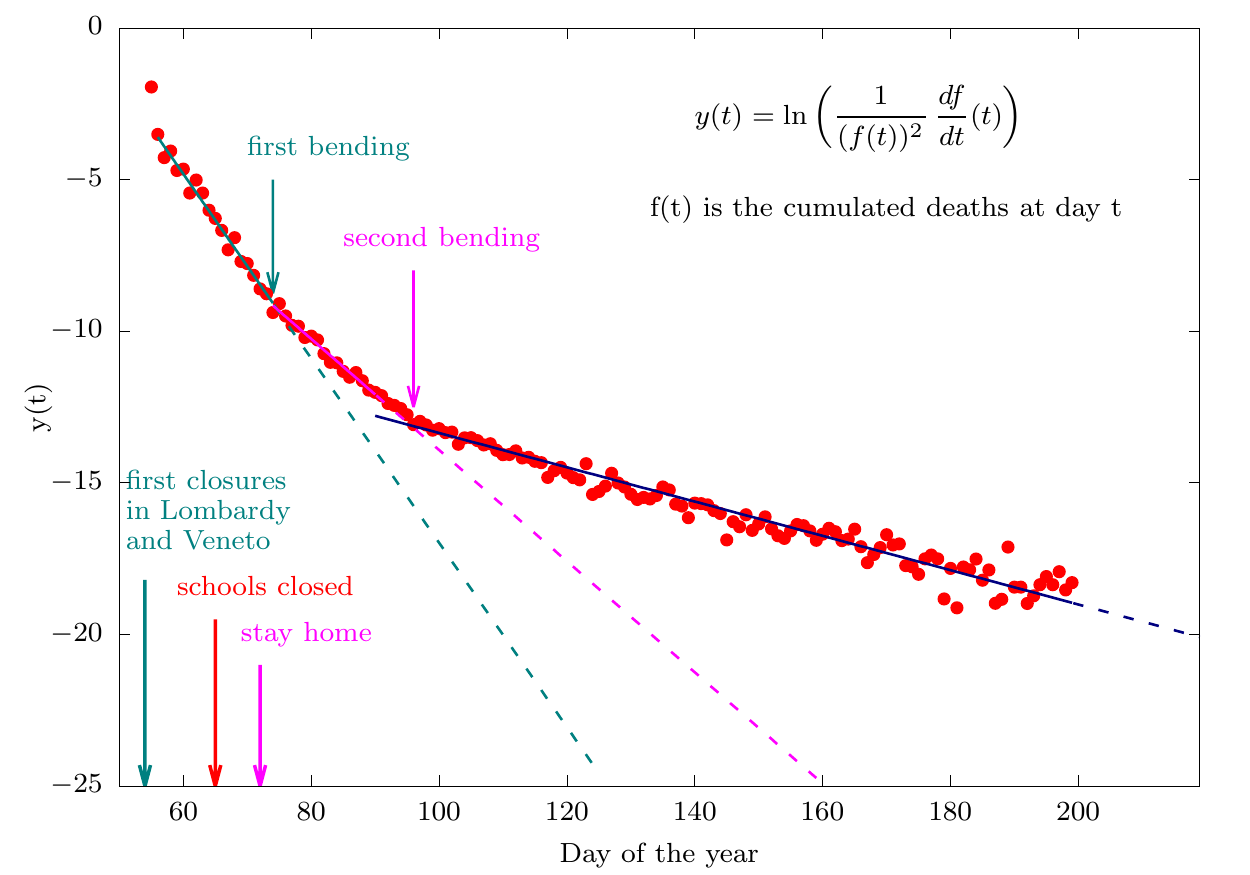}
  \caption{}
  \label{Fig:log_data}
\end{figure}
Fig.\ref{Fig:log_data} refers to the data of the deceased individuals during the winter-spring 2020 {\em first wave} of Covid-19 in Italy: it remarkably confirms this model feature. 
One important point here is that the slopes of the straight segments, that are inversely proportional to the related time constants $\tau$, are completely determined by the parameters $\alpha$ and $\beta$ (besides the initial conditions) and so is the angle between such straight segments: consequently one can compare that angle with the theoretically predicted one and argue about the effects of social measures to reduce the pandemic, of course within the trustworthiness of the model.

\section{Conclusions}
In this paper the equations of the  SIR(D) epidemiological model are replaced by approximate ones, whose solutions are totally defined uniquely by the basic reproduction ratio $\alpha$ and the fractional removal rate $\gamma$ (alternatively by $\beta = \gamma/\alpha$). These solutions are continuous (with the first derivative) chains of two or three or four generalized logistic related functions, the number depending on the value of $\alpha$ only; they are summarized in eq.s \ref{w(t)}\, and easily implementable and usable, for instance, to fit-study data.\newline
The analytic geometry based approximation method used here is novel and set stable at least over the range of the measured values of the basic reproduction ratio for several known pandemic diseases. A useful feature of the SIR(D) model, never disclosed before, is also given. \newline

%



\begin{thebibliography}{00}

\bibitem{Kermack_McKendrick}
  W.~O. Kermack, A.~G. McKendrick, Contribution to the mathematical theory of
  epidemics, Proc. Roy. Soc. A 115 (1927) 700--721.
  \newblock \href {http://dx.doi.org/10.1098/rspa.1927.0118}
  {\path{doi:10.1098/rspa.1927.0118}}.

\bibitem{Murray:1993}
  J.~Murray, Mathematical Biology, Berlin: Springer-Verlag, 1993.
  
\bibitem{Daley_Gani}
  D.~Daley, J.~Gani, Epidemic Modelling, Cambridge University Press, 1999.
  
\bibitem{Brauer:2017}
  F.~Brauer, Mathematical epidemiology: Past, present, and future, Infect Dis
  Model 2 (2017) 113--127.
  \newblock \href {http://dx.doi.org/10.1016/j.idm.2017.02.001}
  {\path{doi:10.1016/j.idm.2017.02.001}}.

\bibitem{Martcheva}
  M.~Martcheva, An Introduction to Mathematical Epidemiology, Springer, 2015.

\bibitem{Brauer_Castillo-Chavez_Feng}
  F.~Brauer, C.~Castillo-Chavez, Z.~Feng, Mathematical Models in Epidemiology,
  Springer, 2019.

\bibitem{Heesterbeek}
  J.~A.~P.~Heesterbeek, A brief history of $R_0$ and a recipe for its calculation,
  DOI: 10.1023/A:1016599411804

\bibitem{Ozyapici Bilgeha}
  A.~Ozyapici, B.~Bilgehan,
Generalized system of trial equation methods and their applications to biological systems,
Applied Mathematics and Computation,
Volume 338,
2018,
Pages 722-732,
ISSN 0096-3003,
https://doi.org/10.1016/j.amc.2018.06.020.

\bibitem{Steven Weinstein}
  N.~S.~Barlow, S.~J.~Weinstein,
Accurate closed-form solution of the SIR epidemic model,
Physica D: Nonlinear Phenomena,
Volume 408,
2020,
132540,
ISSN 0167-2789,
https://doi.org/10.1016/j.physd.2020.132540.

\bibitem{Kroeger Schlickeiser}
  M~Kroeger and R~Schlickeiser, Analytical solution of the SIR-model for the temporal evolution of epidemics. Part A: time-independent reproduction factor, J. Phys. A: Math. Theor. 53 505601, 2020, DOI: 10.1088/1751-8121/abc65d

\bibitem{Pakes}
  A.~G.~Pakes,
  Lambert's W meets Kermack–McKendrick Epidemics,
IMA Journal of Applied Mathematics, Volume 80, Issue 5, October 2015, Pages 1368–1386, %
DOI: 10.1093/imamat/hxu057

\bibitem{Fowler Hollingsworth}
  A~C~Fowler and T~D~Hollingsworth
  Simple approximations for epidemics with exponential and fixed infectious periods,
 Bulletin of mathematical biology, 2015, Springer
 DOI: 10.1007/s11538-015-0095-3 

\bibitem{Cramer}
J.~S.~Cramer, The Origins of Logistic Regression, TI 2002-119/4, Tinbergen Institute Discussion Paper.

  
\end{thebibliography}


\end{document}